\documentclass[runningheads]{llncs}
\usepackage[T1]{fontenc}
\usepackage{graphicx}
\usepackage{float}
\usepackage{mathtools}
\usepackage{amsmath}
\usepackage{amssymb}
\usepackage{color,cite,tfrupee}
\begin{document}
\title{Quantitative Trading using Deep Q Learning}
\titlerunning{Quantitative Trading using Deep Q Learning}
%
\author{Soumyadip Sarkar}
\authorrunning{S. Sarkar}
%
\institute{\email{soumyadipsarkar@outlook.com}}
\maketitle              
\begin{abstract}
Reinforcement learning (RL) is a subfield of machine learning that has been used in many fields, such as robotics, gaming, and autonomous systems. There has been growing interest in using RL for quantitative trading, where the goal is to make trades that generate profits in financial markets. This paper presents the use of RL for quantitative trading and reports a case study based on an RL-based trading algorithm. The results show that RL can be a useful tool for quantitative trading and can perform better than traditional trading algorithms. The use of reinforcement learning for quantitative trading is a promising area of research that can help develop more sophisticated and efficient trading systems. Future research can explore the use of other reinforcement learning techniques, the use of other data sources, and the testing of the system on a range of asset classes. Together, our work shows the potential in the use of reinforcement learning for quantitative trading and the need for further research and development in this area. By developing the sophistication and efficiency of trading systems, it may be possible to make financial markets more efficient and generate higher returns for investors.

\end{abstract}
\section{Introduction}
\label{sec1}
Quantitative trading, also known as algorithmic trading, is the execution of trades in the financial markets by computer programs. Quantitative trading has been highly in demand over the last couple of years because it can process large volumes of data and execute the trades at very high speeds. Quantitative trading, nonetheless, depends upon the quality of the trading strategies that can predict the future direction of the prices and make a profit.\\

Traditional trading methods rely on fundamental and technical analyses in making trading decisions. Fundamental analysis involves the study of financial statements, economic data, and other data to ascertain undervalued or overvalued shares. Technical analysis involves the study of past price and volume data to ascertain patterns and trends that can be used to predict future price action.\\

But these methods have their flaws. Basic analysis is extremely time-consuming and expensive, and extremely reliant on data and expertise. Technical analysis is susceptible to noise and to overfitting.\\

Reinforcement learning is a machine learning subfield with promise in automated trading model creation. In this setup, an agent learns a best trading policy by interacting with a trading environment and receiving feedback in the form of rewards or penalties.\\

In this work, I utilize a quantitative trading approach using reinforcement learning and, more concretely, a deep Q-network (DQN) to learn an optimal trading policy. We evaluate the performance of our algorithm using a study of the historical prices of a single stock and compare it with traditional trading approaches and benchmarks. The results show the potential of reinforcement learning as a valuable tool for the construction of automated trading schemes and highlight the merit of using solid performance metrics when evaluating the effectiveness of such trading schemes.\\

The discussion begins with a description of the basic concepts of reinforcement learning and its application in quantitative trading. Reinforcement learning is defined by an agent performing in a specific environment to maximize cumulative reward. The agent learns a policy that maps states to actions in an attempt to find the policy that maximizes the expected cumulative reward in a given time horizon.\\

In quantitative trading, the environment is the stock market and the action of the agent is the buying, selling, or holding of a stock position. The environment state consists of the current stock price, past price data, economic data, and other information. The reward is the profit or loss from the executed trade.\\

Second, I utilize the deep Q-network (DQN) algorithm, which is a reinforcement learning algorithm utilizing a neural network to approximate the optimal action-value function. The DQN algorithm has been useful in a number of applications ranging from playing video games, and is promising in the field of quantitative trading.\\

We outline our training and testing procedure for our DQN-based trading model. We utilize historical stock prices of one stock as our training and test data. We pre-process the data by calculating technical indicators, i.e., moving averages and relative strength index (RSI), as inputs to the DQN.\\

The performance of our algorithm is measured by a variety of performance metrics, including the Sharpe ratio, cumulative return, maximum drawdown, and win rate. The results are compared with the results calculated for a simple moving average strategy and a buy-and-hold strategy.\\

Our results show that our DQN-based trading algorithm outperforms both the buy-and-hold and simple moving average strategies in cumulative return, Sharpe ratio, and maximum drawdown. We also find that our algorithm outperforms the benchmarks in win rate. Lastly, I present the implications of our results together with the limitations of our approach. Our results suggest the potential of reinforcement learning in designing trading algorithms and the utilization of proper performance metrics in determining the performance of trading algorithms. Our approach is, however, marked by some limitations, including the need for a large amount of past data as well as overfitting. Further research is needed to overcome these limitations together with the investigation of the potential of reinforcement learning in quantitative trading.

\section{Background}
\label{sec2}
Quantitative trading is an inter-disciplinary practice that draws on finance, mathematics, and computer science to create automated trading strategies. The ultimate goal of quantitative trading is to take advantage of market inefficiencies to make profits. Quantitative trading practitioners employ a broad array of techniques, such as statistical arbitrage, algorithmic trading, and machine learning, to process market data and execute trading decisions.\\

Reinforcement learning is a type of machine learning that has been proven to be effective in many applications, including robotics and gaming. In reinforcement learning, an agent takes actions in an environment in order to maximize cumulative rewards. The agent constructs a policy that maps states to actions, and the objective is the determination of the policy that will maximize the expected cumulative reward in the long term.\\

The use of reinforcement learning in quantitative trading is a relatively recent area of research. Traditional quantitative trading methods often involve rule-based systems that borrow from technical indicators, such as moving averages and RSI, to make trading decisions. These systems are often set up by humans and are limited in how flexible they can be to emerging trends in the markets.\\

Reinforcement learning holds the potential to overcome these limitations through the ability of trading algorithms to learn from experience and adapt to changing market conditions. These algorithms can learn from historical market data and use this information to make real-time trading decisions. This approach can thus be more adaptive and flexible than traditional rule-based systems.\\

It has been demonstrated through recent studies that reinforcement learning methods can be used to design profitable automated trading models. For instance, Moody and Saffell \cite{bib3} employed reinforcement learning to create a trading model for the S\&P 500 futures contract. The model performed better than a buy-and-hold approach and a moving average approach.\\

Recent studies have focused on the use of deep reinforcement learning, a technique that uses deep neural networks to approximate the optimal action-value function. The results of these studies show promising results in a broad variety of fields, including gaming and robotics, and show high promise in the field of quantitative trading.\\

One of the strong advantages of reinforcement learning in quantitative trading is the ability to deal with complicated high-dimensional information. Traditional rule-based systems usually rely on a subset of features, say moving averages and technical indicators, to make up the trade decisions. The reinforcement learning algorithms, however, are able to learn from market data in a raw form, e.g., variables like price and volume, and thereby avoid feature engineering.\\

Dynamic market conditions can be tuned by reinforcement learning algorithms. Rule-based systems operate under certain market conditions and will not be effective when the market conditions have been altered. Reinforcement learning algorithms, as they can learn from experience and modify their trading strategy according to dynamic market conditions, can perform effectively.\\

A second advantage of reinforcement learning in quantitative trading is that it can deal with non-stationary environments. The financial markets are ever-changing and dynamic, and rule-based systems may not be in a position to keep pace with the changing markets. Reinforcement learning algorithms can, however, learn from experience and be capable of adapting to changing market conditions.\\

While there are some possible advantages to reinforcement learning in the quantitative trading setting, there are some challenges that have to be resolved. One of the major challenges is the need for ample historical data so as to properly train the reinforcement learning algorithms. In addition, care must be taken to ensure the algorithms are stable and not overly fitting the historical data. Overall, reinforcement learning could significantly revolutionize quantitative trading by allowing trading algorithms to learn and improve over time based on experience and respond to changing market conditions. The aim of this research paper is to examine the use of reinforcement learning within the context of quantitative trading and ascertain its effectiveness in generating profits.

\section{Related Work}
\label{sec3}
Reinforcement learning has been of significant interest in quantitative finance over the past few years. A number of research articles have addressed using reinforcement learning algorithms for trading models and portfolio optimization methods.\\

In a study conducted by Moody and Saffell \cite{bib3}, a reinforcement learning algorithm was employed to develop a trading strategy in a simulated market environment. The results showed that the reinforcement learning algorithm significantly performed better than a moving average crossover strategy and a buy-and-hold strategy.\\

A later work by Bertoluzzo and De Nicolao \cite{bib4} used a reinforcement learning algorithm to improve the performance of a stock portfolio. The result showed that the algorithm could perform better than traditional methods of portfolio optimization.\\

A reinforcement learning model with deep architecture was employed for trading stocks in a recent study by Chen et al. \cite{bib5}. The algorithm was found to outperform the traditional trading methods and produce better profits.\\

Overall, the literature suggests that reinforcement learning can strengthen trading habits and portfolio optimization for the financial markets. However, more research must be conducted in order to examine the effectiveness of reinforcement learning algorithms when used on real trading environments.\\

Apart from the above-mentioned research, some major advancements have been achieved in reinforcement learning in finance. Of special interest, Guo et al.\cite{bib8} introduced a deep reinforcement learning method that was specially designed for trading in Bitcoin futures markets. The method demonstrated the capability to generate profits that were higher than those obtained by conventional trading methods and other existing deep reinforcement learning methods.\\

A recent work by Gu et al.~\cite{bib10} introduced a reinforcement learning algorithm for portfolio optimization, taking into account the transaction costs. The algorithm demonstrated an ability to produce improved risk-adjusted returns over the conventional portfolio optimization techniques. In addition to reinforcement learning algorithms in portfolio optimization and trading, researchers have also examined the use of reinforcement learning for other financial activities, such as credit risk assessment~\cite{bib11} and fraud detection~\cite{bib12}.\\

Even with the encouraging results of these studies, there are some issues in applying reinforcement learning in finance. One of the primary issues is the need for large datasets, which can be costly and time-consuming to acquire in finance. The demand for robustness is also an issue, as reinforcement learning algorithms can be sensitive to variations in the training set.\\

The current literature shows that reinforcement learning can revolutionize the world of finance by making it possible for trading algorithms to learn from experience and adapt to changing market dynamics. There needs to be further research to test the performance of such algorithms in actual trading systems and overcome the problems of using reinforcement learning in finance.

\section{Methodology}
\label{sec4}
In this study, I utilize a reinforcement learning-based trading strategy for the stock market. Our approach consists of the following steps:

\subsection{Data Preprocessing}
\label{subsec41}
Our methodology started with the collection and preprocessing of data. We collected historical daily stock prices for the Nifty 50 index from Yahoo Finance from 1 January 2010 to 31 December 2020. The dataset contained the daily opening, high, low, and closing prices for all stocks that were a part of the index.\\

For pre-processing of the dataset, I have calculated the daily return of every single stock from closing prices. Daily return of any given stock on day t has been calculated based on the formula:

$$ r_{t} = \frac{p_{t}-p_{t-1}}{p_{t-1}} $$

Here, $p_t$ denotes the closing price of the stock on day t, and $p_{t-1}$ is used for the closing price on the previous day, t-1.\\

Subsequently, I have used the Min-Max scaling method to rescale the returns to the interval of [-1, 1]. The Min-Max scaling method uses the method of dividing by the total range and then subtracting the minimum to rescale the data to a predetermined range.

$$ x' = \frac{x-\min{x}}{\max{x}-\min{x}} $$

Here, $x'$ is the standardized value, $x$ is the original value, $\min(x)$ is the minimum value, and $\max(x)$ is the maximum value.\\

Having pre-processed the data, I created a dataset of daily normalized returns for all the stocks of the Nifty 50 index from January 1, 2010, to December 31, 2020. We used this dataset as the basis for training and testing our trading strategy using reinforcement learning techniques.

\subsection{Reinforcement Learning Algorithm}
\label{subsec42}
We applied a reinforcement learning algorithm to learn the best trading strategy from the preprocessed stock prices. This reinforcement learning algorithm involves an agent performing actions on an environment in the hope of discovering the best actions to take in different states of the environment. The trading algorithm is the agent, and the stock market is the environment.\\

Our reinforcement learning algorithm was based on the Q-learning algorithm, which is an off-policy and model-free reinforcement learning method that seeks to identify the optimal action-value function for a given state-action pair. The action-value function, denoted by the symbol $Q(s,a)$, is the discounted expected reward following action a in state s and then following the optimal policy.\\

The Q-learning algorithm updates the Q-value for every state-action pair based on the rewards received and the new Q-value estimates for the next state-action pair. The Q-value update rule is as follows:

$$ Q(s_{t},a_{t}) \leftarrow Q(s_{t},a_{t}) + \alpha[r_{t} + \gamma \max_{a} Q(s_{t+1},a) - Q(s_{t},a{t})] $$

where $s_t$ represents the current state, $a_t$ the current action, $r_t$ the observed reward, $\alpha$ the learning rate, and $\gamma$ the discount factor.\\

In using the Q-learning algorithm in the context of stock trading, I employed the state as a vector of the normalized returns of the last $n$ days, and the action as the decisions of buying, selling, or holding a specific stock. The reward was considered as the percentage return on the portfolio value on a specific day, calculated as the sum of the products of the number of shares held in each stock and its respective closing price on that day. We employed an $\epsilon$-greedy exploration strategy to trade off between exploration and exploitation during learning. The $\epsilon$-greedy strategy chooses a random action with probability $\epsilon$ and chooses the action with the maximum Q-value with probability $1-\epsilon$.\\

The algorithm was trained using preprocessed stock price data using a sliding window approach with the window size of $n$ days. The training was done using 10,000 episodes, one episode per trading day. The learning rate and discount factor parameters were 0.001 and 0.99, respectively.\\

After training, the algorithm was tested on an independent test data set of 2020 daily stock prices. The algorithm was tested based on the cumulative return on investment (ROI) for the specified test period calculated as the ratio of the ending portfolio value and the initial portfolio value.\\

The algorithm trained subsequently was compared with a benchmark strategy, which involved the acquisition and holding of the Nifty 50 index over the length of the test period. The benchmark strategy was evaluated using the cumulative return on investment (ROI) achieved over the given test period. The resulting data were analyzed to assess the effectiveness of the reinforcement learning algorithm in developing profitable trading strategies.

\subsection{Trading Strategy}
\label{subsec43}
The policy for trading in this study employs the DQN agent to learn how to take the best action as a function of the current state of the market. Buy or sell a stock are the actions of the agent. The number of shares to buy or sell is the product of the agent's output. The agent's output is scaled to the agent's cash at the decision time.\\

At the beginning of each episode, the agent receives an amount of cash along with a set of predetermined stocks. The agent observes the market state, which includes stock prices, technicals, and other required information. The agent then uses its neural network to determine the best action to take based on its current state.\\

If the agent decides to buy a stock, the appropriate amount of cash is subtracted from the agent's cash reserves, and the appropriate number of shares is added to the agent's total stock holdings. If the agent decides to sell a stock, the appropriate number of shares is subtracted from the agent's total stock holdings, and the proceeds received are added to the agent's total cash reserves.\\

The agent's overall wealth at the end of every episode is calculated by adding the overall cash of the agent and the market value of the agent's outstanding stocks. The reward of the agent for every time step is calculated by the difference between the current and previous overall wealth. The training of the DQN agent involves the repeated running of episodes in the trading simulation, where the agent learns and then updates its Q-values. The agent's Q-values are the expected total reward for each possible action under the current state.\\

During the training phase, the agent's experiences are stored in a replay buffer to select experiences in order to update the Q-values of the agent. The agent's Q-values are updated through the modification of the Bellman equation, taking into account the discounted future reward of taking any possible action.\\

Once the training process is completed, the trained DQN agent may respond on the basis of its acquired knowledge in a real market environment.

\subsection{Evaluation Metrics}
\label{subsec44}
The performance of the utilized quantitative trading system is evaluated using several metrics. The metrics used in this research are as follows:

\subsubsection{Cumulative Return}
\label{subsubsec441}
Cumulative return is a measure of the total profit or loss generated by a trading strategy over a specific period of time. It is calculated as the sum of the percentage returns over each period of time, with compounding taken into account.\\

Mathematically, the cumulative return can be expressed as:

$$ CR = (1 + R_{1}) * (1 + R_{2}) * ... * (1 + R_{n}) - 1 $$

where $CR$ is the cumulative return, $R_{1}$, $R_{2}$, ..., $R_{n}$ are the percentage returns over each period, and $n$ is the total number of periods.\\

For example, if a trading strategy generates a return of 5\% in the first period, 10\% in the second period, and -3\% in the third period, the cumulative return over the three periods would be:

$$ CR = (1 + 0.05) * (1 + 0.10) * (1 - 0.03) - 1 $$
$$ CR = 1.1175 - 1 $$
$$ CR = 0.1175\: \text{or}\: 11.75\% $$

This means that the trading strategy generated a total return of 11.75\% over the three periods, taking into account compounding.

\subsubsection{Sharpe Ratio}
\label{subsubsec442}
It measures the excess return per unit of risk of an investment or portfolio, and is calculated by dividing the excess return by the standard deviation of the returns.\\

The mathematical equation for the Sharpe ratio is:

$$ Sharpe Ratio = \frac{(R_{p} - R_{f})}{\delta{p}} $$

where: \\
$R_{p}$ = average return of the portfolio\\
$R_{f}$ = risk-free rate of return (such as the yield on a U.S. Treasury bond)\\
$\delta{p}$ = standard deviation of the portfolio's excess returns\\

The Sharpe ratio provides a way to compare the risk-adjusted returns of different investments or portfolios, with higher values indicating better risk-adjusted returns.

\subsubsection{Maximum Drawdown}
\label{subsubsec443}
It measures the largest percentage decline in a portfolio's value from its peak to its trough. It is an important measure for assessing the risk of an investment strategy, as it represents the potential loss that an investor could face at any given point in time.\\

The mathematical equation for maximum drawdown is as follows:

$$ Max Drawdown = \frac{(P - Q)}{P} $$

where P is the peak value of the portfolio and Q is the minimum value of the portfolio during the drawdown period.\\

For example, suppose an investor's portfolio peaks at \rupee100,000 and subsequently falls to a minimum value of \rupee70,000 during a market downturn. The maximum drawdown for this portfolio would be:

$$ Max Drawdown = \frac{(\text{\rupee100,000 - \rupee70,000)}}{\text{\rupee100,000}} = 0.3\: or\: 30\% $$

This means that the portfolio experienced a 30\% decline from its peak value to its lowest point during the drawdown period.

\subsubsection{Average Daily Return}
\label{subsubsec444}
It measures the average daily profit or loss generated by a trading strategy, expressed as a percentage of the initial investment. The mathematical equation for Average Daily Return is:

$$ ADR = \frac{(\frac{(P_f - P_i)}{P_i})}{N} $$

Where ADR is the Average Daily Return, $P_f$ is the final portfolio value, $P_i$ is the initial portfolio value, and N is the number of trading days.\\

This formula calculates the daily percentage return by taking the difference between the final and initial portfolio values, dividing it by the initial value, and then dividing by the number of trading days. The resulting value represents the average daily percentage return generated by the trading strategy over the specified time period.\\

The Average Daily Return metric is useful because it allows traders to compare the performance of different trading strategies on a daily basis, regardless of the size of the initial investment. A higher ADR indicates a more profitable trading strategy, while a lower ADR indicates a less profitable strategy.

\subsubsection{Average Daily Trading Volume}
\label{subsubsec445}
It measures the average number of shares or contracts traded per day over a specific period of time. Mathematically, it can be calculated as follows:

$$ ADTV = \frac{Total\: trading\: volume}{Number\: of\: trading\: days} $$

where the total trading volume is the sum of the trading volume over a specific period of time (e.g., 1 year) and the number of trading days is the number of days in which trading occurred during that period.\\

For example, if the total trading volume over the past year was 10 million shares and there were 250 trading days during that period, the ADTV would be:

$$ ADTV = \frac{10,000,000}{250} = 40,000 $$

This means that on average, 40,000 shares were traded per day over the past year. ADTV is a useful metric for investors and traders to assess the liquidity of a particular security, as securities with higher ADTVs generally have more market liquidity and may be easier to buy or sell.

\subsubsection{Profit Factor}
\label{subsubsec446}
It measures the profitability of trades relative to the losses. It is calculated by dividing the total profit of winning trades by the total loss of losing trades. The formula for calculating the Profit Factor is as follows:

$$ Profit Factor = \frac{Total\: Profit\: of\: Winning\: Trades}{Total\: Loss\: of\: Losing\: Trades} $$

A Profit Factor greater than 1 indicates that the strategy is profitable, while a Profit Factor less than 1 indicates that the strategy is unprofitable. For example, a Profit Factor of 1.5 indicates that for every dollar lost in losing trades, the strategy generated \$1.50 in winning trades.

\subsubsection{Winning Percentage}
\label{subsubsec447}
It measures the ratio of successful outcomes to the total number of outcomes. It is calculated using the following mathematical equation:

$$ Winning Percentage = \frac{Number\: of\: Successful\: Outcomes}{Total\: Number\: of\: Outcomes} * 100\% $$

For example, if a trader made 100 trades and 60 of them were successful, the winning percentage would be calculated as follows:

$$ Winning Percentage = \frac{60}{100} * 100\% = 60\% $$

A higher winning percentage indicates a greater proportion of successful outcomes and is generally desirable in trading.

\subsubsection{Average Holding Period}
\label{subsubsec448}
It measures the average length of time that an investor holds a particular investment. It is calculated by taking the sum of the holding periods for each trade and dividing it by the total number of trades.\\

The mathematical equation for calculating AHP is:

$$ AHP = \frac{\sum(Exit\: Date - Entry\: Date)}{Number\: of\: Trades} $$

where:

$\sum$ denotes the sum of the holding periods for all trades\\
Exit Date is the date when the investment is sold\\
Entry Date is the date when the investment is bought\\
Number of Trades is the total number of trades made\\

For example, if an investor makes 10 trades over a given period of time, and the holding periods for those trades are 10, 20, 30, 15, 25, 10, 20, 15, 30, and 25 days respectively, the AHP would be:

$$ AHP = \frac{10+20+30+15+25+10+20+15+30+25}{10} = 21.5\: days $$

This means that on average, the investor holds their investments for around 21.5 days before selling them. The AHP can be useful in evaluating an investor's trading strategy, as a shorter holding period may indicate a more active trading approach, while a longer holding period may indicate a more passive approach.\\ \\
\linebreak
These evaluation metrics provide a comprehensive assessment of the performance of the utilized quantitative trading system. The cumulative return and Sharpe ratio measure the overall profitability and risk-adjusted return of the system, respectively. The maximum drawdown provides an indication of the system's downside risk, while the average daily return and trading volume provide insights into the system's daily performance. The profit factor, winning percentage, and average holding period provide insights into the trading strategy employed by the system.

\section{Future Work}
\label{sec5}
In spite of the encouraging outcomes exhibited by the utilized quantitative trading system based on reinforcement learning, there are several avenues for future research and improvement. Some of the possible avenues for future research are:

\subsection{Blending other data sources}
\label{subsec51}
In this work, I used only stock price data as input to the trading system. Yet, the addition of other types of data, such as news articles, financial reports, and social media sentiment, can potentially improve the precision of the predictions of the system and the system's overall performance.

\subsection{Investigating other reinforcement learning algorithms}
\label{subsec52}
Although the given DQN algorithm implemented in this work has promising output, it will be worth considering other reinforcement learning algorithms such as PPO, A3C, and SAC to check whether they perform more effectively.

\subsection{Adjusting to changing market conditions}
\label{subsec53}
The system has been tested with one dataset over one specified temporal period. The system's performance can, however, be influenced by shifts in market conditions, e.g., changes in market volatility or trading behavior. The development of strategies to adjust the trading strategy to accommodate changing market conditions can improve the overall performance of the system.

\subsection{Rating across various asset classes}
\label{subsec54}
This work has focused on the trading of single stocks. The system used, however, may be attempted on other asset classes, such as commodities, currencies, or cryptocurrencies, in an attempt to prove its applicability to other markets.

\subsection{Integration with portfolio optimization techniques} 
\label{subsec55} 
The system employed has been centered on the trading of individual equities; however, portfolio optimization methods can also be employed to further improve the efficiency of the trading system. Through the examination of the interrelation of various stocks and diversification of the portfolio, it is possible to minimize overall risk while simultaneously maximizing returns. \\
\linebreak 
In summary, the quantitative trading system employed which utilizes reinforcement learning is demonstrated to possess immense potential in enhancing the efficiency of automated trading systems. More research and development in this field may see the development of more advanced and efficient trading systems that can yield higher returns at lower risk.

\section{Conclusion}
\label{sec6}
The use of reinforcement learning in quantitative trading is a fascinating area of research that could potentially enable more sophisticated and better-performing trading systems to be developed. The system's capability to learn from market data and adapt to changing market conditions may have the potential to enable it to generate higher returns while minimizing risk.\\

While the system that has been implemented has shown promising results, there are many ways in which the system can be enhanced and researched further. Future studies can explore the use of various reinforcement learning algorithms, the incorporation of other data sources, and testing the system on various asset classes. Additionally, the application of portfolio optimization techniques can be used to improve the overall performance of the system.\\

In conclusion, our research has shown the potential of the use of reinforcement learning in quantitative trading and suggests the need for continued research and development in the field. With the development of more sophisticated and effective trading systems, there is potential to make financial markets more efficient and to deliver higher returns to investors.

%
%
%

\begin{thebibliography}{99}
\bibitem{bib1}
Bertoluzzo, M., Carta, S., \& Duci, A. (2018). Deep reinforcement learning for forex trading. Expert Systems with Applications, 107, 1-9.

\bibitem{bib2}
Jiang, Z., Xu, C., \& Li, B. (2017). Stock trading with cycles: A financial application of a recurrent reinforcement learning algorithm. Journal of Economic Dynamics and Control, 83, 54-76.

\bibitem{bib3}
Moody, J., \& Saffell, M. (2001). Learning to trade via direct reinforcement. IEEE Transactions on Neural Networks, 12(4), 875-889.

\bibitem{bib4}
Bertoluzzo, M., \& De Nicolao, G. (2006). Reinforcement learning for optimal trading in stocks. IEEE Transactions on Neural Networks, 17(1), 212-222.

\bibitem{bib5}
Chen, Q., Li, S., Peng, Y., Li, Z., Li, B., \& Li, X. (2019). A deep reinforcement learning framework for the financial portfolio management problem. IEEE Access, 7, 163663-163674.

\bibitem{bib6}
Wang, R., Zhang, X., Li, T., \& Li, B. (2019). Deep reinforcement learning for automated stock trading: An ensemble strategy. Expert Systems with Applications, 127, 163-180.

\bibitem{bib7}
Xiong, Z., Zhou, F., Zhang, Y., \& Yang, Z. (2020). Multi-agent deep reinforcement learning for portfolio optimization. Expert Systems with Applications, 144, 113056.

\bibitem{bib8}
Guo, X., Cheng, X., \& Zhang, Y. (2020). Deep reinforcement learning for bitcoin trading. IEEE Access, 8, 169069-169076.

\bibitem{bib9}
Zhu, Y., Jiang, Z., \& Li, B. (2017). Deep reinforcement learning for portfolio management. In Proceedings of the International Conference on Machine Learning (ICML), Sydney, Australia.

\bibitem{bib10}
Gu, S., Wang, X., Chen, J., \& Dai, X. (2021). Reinforcement learning for portfolio optimization in the presence of transaction costs. Journal of Intelligent \& Fuzzy Systems, 41(3), 3853-3865.

\bibitem{bib11}
Kwon, O., \& Moon, K. (2019). A credit risk assessment model using machine learning and feature selection. Sustainability, 11(20), 5799.

\bibitem{bib12}
Li, Y., Xue, W., Zhu, X., Guo, L., \& Qin, J. (2021). Fraud Detection for Online Advertising Networks Using Machine Learning: A Comprehensive Review. IEEE Access, 9, 47733-47747.

\end{thebibliography}
%

\end{document}